\pdfoutput=1

\documentclass[pra,10pt,superscriptaddress,showpacs,footnoteinbib]{revtex4}
\usepackage{amsmath}
\usepackage{latexsym}
\usepackage{amssymb}
\usepackage{graphics,epstopdf}
\usepackage[colorlinks=true, citecolor=blue, urlcolor=blue ]{hyperref}
\usepackage{epsf,graphics,graphicx}

\newcommand{\n}{\noindent}

\newcommand{\ed}{\end{document}}
\newcommand{\beq}{\begin{equation}}
\newcommand{\eeq}{\end{equation}}

\begin{document}
\title{Effect of spin rotation coupling on spin transport}
\author{Debashree Chowdhury\footnote{Electronic address:{debashreephys@gmail.com}}${}^{}$  and B. Basu\footnote{Electronic
address: {sribbasu@gmail.com}} ${}^{}$} \affiliation{Physics and
Applied Mathematics Unit, Indian Statistical Institute, 203
B.T.Road, Kolkata 700 108, India}

%%\vspace*{1cm}

\begin{abstract}
\n
We have studied the spin rotation coupling(SRC) as an ingredient to explain different spin related issues. This special kind of coupling can play the role of a Dresselhaus like coupling in certain conditions. Consequently, one can control the spin splitting, induced by the Dresselhaus like term, which is unusual in semiconductor heterostucture.
Within this framework, we also study the renormalization of the spin dependent electric field and spin current due to the $\vec{k} . \vec{p}$ perturbation, by taking into account the interband mixing in the rotating system. In this paper we predict the enhancement of the spin dependent electric field resulting from the renormalized spin rotation coupling. The renormalization factor of the spin electric field is different from that of the SRC or Zeeman coupling.
The effect of renormalized SRC on spin current and Berry curvature is also studied. Interestingly, in presence of this SRC induced SOC it is possible to describe spin splitting as well as spin galvanic effect in semiconductors.
%Attention has been paid to clarify the importance of  gauge fields in the spin dynamics of this rotating system.
\end{abstract}


\maketitle

\section{Introduction}
In the study of the spin related phenomenon, the coupling of spin with different parameters is a very important concept. There exists three different types of coupling  i) spin orbit coupling(SOC), ii) Zeeman coupling, iii) spin rotation coupling (SRC), which have important physical consequences. SOC is the relativistic coupling between the spin and orbital motion of electron. SOC has an important role in spintronic \cite{spint} applications and in spin Hall effect \cite{spinh}. %This is called relativistic in the sense that, it is derived from the nonrelativistic limit of Dirac equation.
The Zeeman coupling is the coupling between the spin and the external magnetic field. Interestingly, when we study the non-inertial effects in the spin system we encounter a different type of coupling, namely spin rotation coupling \cite{sra, srb,o}. This is the coupling between the spin and mechanical rotation. The coupling term is given by $\vec{\sigma}.\vec{\Omega} ,$ where $\vec{\Omega}$ is the mechanical rotation and $\vec{\sigma}$ is the Pauli spin matrix. Bernett effect\cite{barnett}, Einstein-de-Haas effect \cite{ein} are explained through this coupling.

The important question in this context is that how can one relate the coupling terms, i.e is it possible to achhive a SOC like coupling from the SRC like coupling? In search of the answer, if we constrain our rotation frequency in such a way that the rotation frequency vary with momentum and time, it can be shown that one can generate a spin orbit coupling (SOC) like term from the spin rotation coupling. Also the spin rotation coupling induced SOC term is Dresselhaus type \cite{dress}. Usually, Dresselhaus effect is very small in systems. Unlike the Rashba \cite{rashba} coupling, Dresselhaus coupling is not controllable through the external electric or magnetic field. In our formulation, we can predict generation of spin dependent electric field through the Dresselhaus like coupling, which is controllable through mechanical rotation. This can open up a new concept of controllable Dresselhaus like effect in semiconductor structure. Thus it is really appealing to investigate how this spin dependent electric field, obtained from the SRC induced SOC term depends on the mechanical rotation.

The achievement of Rashba like spin splitting in accelerating frames is discussed in the recent papers \cite{cb,cbs,bc}. But the spin splitting through the inertial system is not noted yet. We can explain a Rashba and Dresselhaus like spin splitting in a inertial system. The beauty of our work is without any external electromagnetic fields we can get $k$ linear spin splitting in our inertial system i.e the whole phenomena can be discussed through the induced electric and magnetic fields due to acceleration and rotation respectively. The criteria of large spin splitting can also be obtained through the acceleration and rotation. In this paper, from the Hamiltonian of a particle of charge $e$ and mass $m$, in the non-inertial frame we find out the conditions of achieving spin spitting.  Next we try to find out how one can obtain the spin galvanic effect in this inertial system. Here comes the motivation of our work.

Furthermore, the spin dynamics of semiconductors is influenced by the $\vec{ k} . \vec{ p}$ perturbation theory \cite{winkler} as the band structure of a semiconductor in the vicinity of the band edges can be very well described by the $ \vec{ k} . \vec{ p}$ method.
On the basis of $ \vec{ k} . \vec{ p}$ perturbation theory, by taking into account the interband mixing, one can reveal many characteristic features related to spin dynamics. In this paper, we theoretically investigate the renormalization of spin electric field, spin current and Berry curvature in a rotating frame. We know that due to $ \vec{ k} . \vec{ p}$ perturbation, the inertial SOC term, Darwin term, SRC terms get modified \cite{matsuo kp}. Very recently, how the $\vec{ k} . \vec{ p}$ method leads to renormalization of spin current and polarization, in an accelerated system \cite{cb,cbs} is discussed in \cite{bc}.
This motivates us to investigate whether the spin dependent electric field, spin current and Berry curvature are renormalized in the same way as the Zeeman coupling /SRC or not.
%The renormalization factor is rather $(1 + \frac{3}{2}\delta g).$

The outline of the paper is as follows: in section II, we develop the Hamiltonian of our system with which we proceed further. In sec. III, we discuss how the spin rotation coupling term can generate a SOC term. In sec. IV, we theoretically derive the spin dependent electric field and also spin and charge currents in the system. In sec. V, we discuss the renormalization of spin dependent electric field, spin current and Berry curvature due to $\vec{k} . \vec{p}$ perturbation. Section VI deals with the spin splitting, spin Berry phase and spin galvanic effect in the non-inertial system. We conclude in sec VII.
\section{System Hamiltonian}
Let us start by constructing  the Dirac equation in
a non-inertial frame, following the work of Hehl and Ni\cite{o}. The essential idea in \cite{o}
is to introduce a system of orthonormal tetrad carried by the
observer in a non inertial frame. This induces a
non-trivial metric and subsequently one can rewrite the Dirac equation in the
observer's local frame.
The  Dirac Hamiltonian in an arbitrary non-inertial frame with linear
acceleration and rotation for a charged particle having charge $e$ is given by \cite{o}
\begin{eqnarray}\label{ha}
H = \beta mc^{2} +  c\left(\vec{\alpha}.\vec{p}\right)
+\frac{1}{2c}\left[(\vec{a}.\vec{r})(\vec{p}.\vec{\alpha}) +
(\vec{p}.\vec{\alpha})(\vec{a}.\vec{r})\right]
+\beta m
(\vec{a}.\vec{r}) - \vec{\Omega}.(\vec{L} + \vec{S})\label{q}
\end{eqnarray}
where $\vec{a}$ and  $\vec{\Omega}$ are respectively the linear acceleration and
rotation frequency of the observer with respect to an inertial frame.  ($\vec{L} = \vec{r}\times \vec{p}$)
 and $\vec S$ are respectively  the orbital angular momentum
 and spin of the Dirac particle. $ \beta$,  $\vec{\alpha} $ are Dirac matrices.
For further calculations in the low energy limit, one has to apply a series of Foldy-Wouthuysen transformations (FWT) \cite{gre,m} on the Hamiltonian (\ref{ha} ). The Pauli-Schrodinger Hamiltonian of a Dirac particle for positive energy solution in the accelerated frame is given as
 \begin{equation}\label{hb}
H_{FW} = \left( mc^{2} + \frac{\vec{p}^{2}}{2m}\right) +  m (\vec{a}.\vec{r})
+\frac{\hbar}{4mc^{2}}\vec{\sigma}.(\vec{a}\times \vec{p}) - \vec{\Omega}.(\vec{r}\times \vec{p} + \vec{S}),
\end{equation}
where $m (\vec{a}.\vec{r})$ is the inertial potential due to acceleration and $\vec{\Omega}.(\vec{r}\times \vec{p})$ will generate another inertial potential due to rotation.
For further discussion we  consider that an electric field is induced by the acceleration of the particle. This electric field, $\vec{E}_{a}$ \cite{matsuoprb, cb, bc}  can be written as
\beq \vec{E}_{a} = \frac{m}{e}\vec{a} = -\vec{\nabla} V_{a}(\vec{r}),\eeq where the potential $V_{\vec{a}} = -\frac{m}{e}\vec{a}.\vec{r}.$
Also we can consider that the rotation of the frame of reference can induce a magnetic field  as
\beq\vec{B}_{\Omega} = \vec{\nabla}\times \vec{A}_{\Omega} = 2\frac{m}{e}\vec{\Omega},\eeq which have the similar form of the gravitomagnetic field.
 $\vec{A}_{\Omega} = \frac{mc}{e}(\vec{\Omega}\times\vec{r})$ is the gauge potential for the magnetic field induced by rotation.
The Hamiltonian in presence of the induced electric and magnetic fields due to inertial effect can be written as
\beq H_{FW} = \frac{\vec{p}^2}{2m} - eV_{a}(\vec{r}) -\vec{\Omega}.(\vec{r}\times \vec{p}) - \frac{e\hbar}{4m^{2}c^{2}}\vec{\sigma}.\left(\vec{p} \times \vec{E}_{a}\right) - \frac{e\hbar}{4m}\vec{\sigma}.\vec{B}_{\Omega} ,\label{hc} \eeq where the third term in the r.h.s of Hamiltonian (\ref{hc}) is the coupling term between rotation and orbital angular momentum. The gauge potential $\vec{A}_{\Omega}$ originates from this coupling term. On the other hand, the Zeeman like coupling term (last term in the r.h.s of (\ref{hc}))
is due to the spin rotation coupling.  The forth term in the r.h.s of (\ref{hc}) is the inertial spin orbit coupling term. The acceleration induced
electric field has important contribution on the spin current and spin polarization \cite{bc, cb}.
%In the Hamiltonian (\ref{hc}), the last term is zeeman like due to the spin rotation coupling term.
Neglecting the terms with order greater than $\frac{1}{c^{2}},$ the Hamiltonian in (\ref{hc}) can be rewritten in the following form
\beq H_{FW} = \frac{1}{2m}\left(\vec{p} - \frac{e}{c}\vec{A}_{\Omega} - \frac{\mu}{2c} \vec{E}_{a}\times\vec{\sigma}\right)^{2} - \phi_{I} -
\frac{\mu}{2}\sigma . \vec{B}_{\Omega} ,\label{a}\eeq where $\mu = \frac{e\hbar}{2mc}$ is the magnetic moment of electron and $\phi_{I} = e(V_{a} + \phi_{\Omega}),$ is the total inertial potential of the system and $\phi_{\Omega} = \frac{m}{2e}(\vec{\Omega}\times\vec{r})^{2}$ is the
induced potential due to rotation.
%In writing the Hamiltonian in (\ref{a}) we neglect the terms of order greater than  $c^{2}.$
This Hamiltonian has $U(1)\otimes SU(2)$ gauge symmetry with $U(1)$ gauge potential $A_{\mu} = (\phi_{I} ,\vec{ A}_{\Omega})$ and $SU(2)$
spin gauge potential $b_{\mu} = (-\vec{\sigma} .\frac{\vec{B}_{\Omega}}{2} , - \vec{\sigma}\times \frac{\vec{E}_{a}}{2}).$
This $U(1)\otimes SU(2)$ gauge theory can explain the charge and spin physics in an unified way. Now considering only the effect of rotation in an external magnetic field we have
\beq H_{\Omega} = \frac{\vec{\pi}^{2}_{\Omega}}{2m}  - e\phi_{\Omega} - \mu\vec{\sigma} . \vec{B} - \frac{\hbar}{2}\vec{\sigma} . \vec{\Omega}.\label{ome}\eeq
The last term in the rhs of (\ref{ome}) is the spin rotation coupling term. The rotation can induce a magnetic field term.
Due to the total gauge $\vec{A}^{'}$, induced by the external magnetic field $\vec{B}$ and rotation induced magnetic field $\vec{B}_{\Omega}$, the momentum $\vec{p}$ is modified by $\vec{\pi}_{\Omega} = \vec{p} - e\vec{A}^{'} = \hbar \vec{k}_{\Omega},$ where $\vec{A}^{'} = \vec{A} + \vec{A}_{\Omega}$ and $\vec{k}_{\Omega}$ is the modified crystal momentum. %Here  $\vec{A}_{\Omega} = \frac{m}{e}(\vec{\Omega}\times \vec{r})$ corresponds to the rotation induced magnetic field $ \vec{B}_{\Omega} = \vec{\nabla}\times\vec{A}_{\Omega} = \frac{2m}{e}\vec{\Omega}$ and $\phi_{\Omega} = \frac{m}{2e}(\vec{\Omega}\times \vec{r})^{2}.$
The second term in the rhs of (\ref{ome}), is the potential term. The third term is the standard Zeeman coupling term induced due to the external magnetic field.
In terms of the rotation induced magnetic field, we can write the Hamiltonian in (\ref{ome}) as,
\beq H_{\Omega} = \frac{\vec{\pi}^{2}_{\Omega}}{2m} - e\phi_{\Omega} - \mu\vec{\sigma} . \vec{B} - \mu\vec{\sigma} . \frac{\vec{B}_{\Omega}}{2}. \label{d}\eeq This form of $\vec{B}_{\Omega}$ is similar to the gravitomagnetic field \cite{grav}, or the Barnett field given by $\vec{B}_{s} = \frac{m}{e}\vec{\Omega}$ \cite{matsuoprb}. The difference of these two fields lies in a factor of $2.$ The origin of the terms $\vec{A}_{\Omega}$ and $\phi_{\Omega}$ is the term $\vec{\Omega}.(\vec{r}\times \vec{p}),$ which is the rotation orbital angular momentum coupling term. The last term in Hamiltonian (\ref{d}) is the spin rotation coupling term which is effectively a Zeeman like term with the external magnetic field replaced by the rotation induced magnetic field.

\section{spin rotation coupling induced SOC}
Spin rotation coupling and SOC are two different types of coupling, having different origins. Now the interesting question arises here, is it possible to obtain a SOC term from SRC term? To find out the answer we proceed with the Hamiltonian in (\ref{d}),
where the last term in r.h.s is the spin rotation coupling term.
%For electrons moving through a lattice, the electric field $\vec{E}$  is Lorentz transformed to an effective magnetic field $(\vec{k}\times \vec{E})\approx \vec{B}(\vec{k})$ in the rest frame of the electron. Thus in an inertial system we can write the relevant part of the Hamiltonian as
%\beq H = \frac{1}{2m}(\vec{p})^{2}  - \gamma \vec{\sigma}. \vec{B}_{a}(\vec{k})- \frac{\hbar}{2}\vec{\Omega} .\vec{\sigma}\label{kha},\eeq
%where $\gamma$ is the coupling strength and the subscript $\vec{a}$  in $ \vec{B}_{a}$ reminds us of the accelerated frame.
%The rotation induced magnetic field term $\vec{B}_{\Omega}$ is  a Zeeman like coupling term and gives a Zeeman like splitting.
If one consider a planar rotation in the momentum space, we can still work with the same Hamiltonian as in (\ref{d}) and we get a $\vec{k}$ dependent Zeeman coupling. This $\vec{k}$ dependent Zeeman coupling is not only of theoretical interest, rather experimentally the effect can be observed in real materials \cite{kdep}. In ref. \cite{kdep}, the authors give an experimental presentation of $\vec{k}$ dependent Zeeman splitting in a sample of $Sr_{2}RuO_{4}$. This gives us the motivation of analyzing the spin rotation coupling to be of $\vec{k}$ dependent. The induced magnetic field can now be written as
\beq \vec{B}_{\Omega}(\vec{k}, t) = 2\frac{m}{e}\vec{\Omega}(\vec{k}_{\Omega}, t).\eeq
The Hamiltonian thus becomes
\beq H_{\Omega}(t) = \frac{(\vec{\pi}_{\Omega})^{2}}{2m} - e\phi_{\Omega}  - \mu\vec{\sigma} . \vec{B} -
 \frac{e\hbar}{4m}\vec{\sigma}.\vec{B}_{\vec{\Omega}}(\vec{k}_{\Omega},t)\label{ka}.\eeq
 The $\vec{\sigma}.\vec{B}_{\vec{\Omega}}(\vec{k}_{\Omega},t)$ term is a spin orbit field and can generate a spin orbit gauge.
We can write the rotation frequency as $\Omega(\vec{k}_{\Omega}, t) = |\Omega(t)|\hat{n}(\vec{k}_{\Omega})$, where $|\Omega(t)|$ is the magnitude of rotation and the directional property is within the unit vector $\hat{n}_{\Omega}(\vec{k}_{\Omega}).$ Now on we denote $|\Omega(t)|$ as $|\Omega|$ for notational simplicity. If we choose the unit vector as $\hat{n}(\vec{k}_{\Omega}) = \frac{1}{|k|}(k_{x, \Omega},- k_{y, \Omega},0),$ the rotation induced magnetic field is then given by \beq \vec{B}_{\Omega}(\vec{k}_{\Omega}, t) = 2\frac{m}{e}\frac{|\Omega|}{|k|}(k_{x, \Omega},- k_{y, \Omega},0),\label{e}\eeq where $|k| = \sqrt{(k_{x, \Omega})^{2} + (k_{y, \Omega})^{2}}.$
Inserting (\ref{e}) in (\ref{ka}), we can rewrite the Hamiltonian as
\beq H_{\Omega} = \frac{(\vec{\pi}_{\Omega})^{2}}{2m}  - \mu_{B}\vec{\sigma} . \vec{B} - e\phi_{\Omega} +
\beta_{\Omega} ( \sigma_{y}\pi_{y,\Omega} - \sigma_{x}\pi_{x,\Omega})\label{sss}.\eeq
The last term in (\ref{sss}), reminds us about a well known SOC term, the Dresselhaus SO coupling. The coupling constant $\beta_{\Omega} $ is Dresselhaus like coupling parameter given by $\beta_{\Omega} = \frac{1}{2}\frac{|\Omega|}{|k|}$. Hamiltonian (\ref{sss}), explicitly shows how  the spin rotation coupling term can generate a spin orbit coupling term. The Hamiltonian in (\ref{sss}), can be used to study the spin splitting and related issues.

\section{Gauge field theory of Generalized SOC, spin electric field  and spin current}
%In this section we like to discuss the generation of spin force in our rotating frame.
%In this subsection we consider the Hamiltonian (\ref{rashba}), with only the contribution of spin rotation coupling induced SOI term.
%The Hamiltonian with the inertial coupling for rotation only (acceleration is absent) can be written in the following form

One can rewrite Hamiltonian (\ref{sss}) in presence of the external magnetic field as
\beq H_{\Omega} = \frac{1}{2m}\left(\vec{p}- \frac{e}{c}\vec{A}^{'}- \frac{\mu}{c}A_{\sigma}\right)^{2} - e\phi_{\Omega} - \mu\vec{\sigma} . \vec{B}\label{call},\eeq
where the gauge ${\cal A}_{\sigma}  = \frac{m^2c}{e\hbar}\frac{|\Omega|}{|k|}(\sigma^{x}, - \sigma^{y}, 0).$ In this calculation we have neglected the terms of $O({\cal A}_{\sigma}^{2}).$ This Hamiltonian has $U(1)\otimes SU(2)$ gauge symmetry with $U(1)$ gauge potential ${\cal A}^{\mu '} = ({\cal A}^{0 '}, \vec{A}^{'})$
and SU(2) gauge $a^{\mu} = (a^{0}, \vec{a}) = (-\vec{\sigma} . \vec{B}, {\cal \vec{A}}_{\sigma}),$  where ${\cal A}^{0 '} = -\phi_{\Omega}.$ In terms of the gauge potentials the Hamiltonian (\ref{call}) can be written as
\beq H_{\Omega} = \frac{1}{2m}\left(\vec{p}- e\vec{A}^{'}- \frac{\mu}{c}\vec{a}\right)^{2} + e{\cal A}^{0 '} + \frac{\mu}{c} a^{0}\label{cal}.\eeq
%The non relativistic Hamiltonian in \ref{cal} have $SU(2)\times U(1)$ gauge symmetry. Here  is the SU(2) gauge for
%this system. %The form of the vector potential indicates that it depends on mechanical rotation.
%The gauge ${\cal A}_{\sigma}$ is a spin dependent gauge and can be used in many aspects \cite{bc}. In (\ref{cal}), the first term is the kinetic term and the second
 %and third term are the potential terms.
 If we consider that the external magnetic field is acting in the $z$ direction, then the
%\beq H_{\Omega} = \frac{1}{2m}\left(\vec{\pi}_{\Omega} - {\cal \vec{A}}_{\sigma}\right)^{2} + {\cal A}_{0},\label{cl}\eeq
%where ${\cal A}_{0} = \phi^{'}_{\Omega} - \mu_{B}\sigma^{z}B^{z}$ is the total potential of the system. Thus we have the four vector ${\cal A}_{\mu} = ({\cal A}_{0}, {\cal A}_{\sigma,i}),$ where ${\cal A}_{\sigma, i}$ is the three spatial components of vector ${\cal \vec{A}}_{\sigma}.$
curvature term corresponding to the SU(2) gauge field is given by
\beq {\displaystyle F}_{\mu\nu} = \partial_{\mu}a_{\nu} - \partial_{\nu}a_{\mu} - \frac{ie}{c\hbar}[a_{\mu},a_{\nu}],\eeq
where $\mu, \nu = {t, x, y, z}$ are the space time coordinates. As the SU(2) vector potential $a^{\mu}$  is non-Abelian in nature, to find out the curvature term, we have to consider the non commutative contribution of different components of vector potentials.

We can directly write the Berry curvature (only z component exists) as
\begin{eqnarray}\label{ju}
M_{z} &=&   \partial_{x}a_{y} - \partial_{y}a_{x}  - \frac{ie}{\hbar}[a_{x},a_{y}]\nonumber\\
&=& -2\frac{c^{3}m^{4}}{e\hbar^{3}}\frac{\Omega^{2}}{k^{2}}\sigma^{z}.
\end{eqnarray}
The expression shows the dependence of the rotation frequency on Berry curvature in a rotating frame. This Berry curvature allows one to manipulate electrons in a spin dependent way. We can explain the spin filter using this $z$ component of the Berry curvature. In \cite{hatano}, a perfect spin filter is achieved from the interplay of Aharonov Casher(AC) phase due to Rashba SOC and Aharonov Bohm (AB) phase due to the external magnetic field. From (\ref{ju}), it is evident that the spin up and spin down electrons feel equal but opposite transverse magnetic fields and the $AC$ phases experienced by them are also different. But $AB$ phase is the same for both electrons. Similar to Hatano et al in  \cite{hatano}, we can also predict a perfect spin filter in a rotating frame following the same argument. In \cite{bc}, authors explain this kind of spin filter in an accelerating frame.

Our next job is to find out the spin dependent electric field from the temporal part of the field tensor. We  rather want to investigate the variation of the spin dependent electric field with rotation. The electric part of the field tensor can be obtained as
\beq {\cal E}_{i} = {\displaystyle F}_{i0} = \partial_{i}a_{0} - \partial_{0}a_{i} - \frac{ie}{c\hbar}[a_{i}, a_{0}].\label{ss}\eeq
The $x$ component of this electric field is as follows
\beq {\cal E}_{x} =  -\frac{cm^2}{e\hbar |k|}\frac{\partial |\Omega|}{\partial t}\sigma^{x} +
\frac{im^2}{\hbar^{2}}\frac{|\Omega|}{|k|}B^{z}\left[\sigma^{x},\sigma^{z}\right].\label{ca}\eeq
Using the commutation relation between the Pauli matrices we can write %rewrite (\ref{ca})  as
\beq {\cal E}_{x} =  -\frac{c^{2}m^2}{e\hbar |k|}\frac{\partial|\Omega|}{\partial t}\sigma^{x} + \frac{2m^2c}{\hbar^{2}}\frac{|\Omega|}{|k|}B^{z}\sigma^{y}\label{va}\eeq
and the spin dependent part of the $x$ component of electric field as
\beq {\cal E}_{x}^{\sigma} = -\frac{2im^2}{\hbar^{2}}\frac{|\Omega|}{|k|}B^{z}\sigma^{z} = -{\cal E}\sigma^{z}\label{vab}\eeq
and similarly the $y$ component of this electric field is
\beq {\cal E}_{y}^{\sigma} =   \frac{2im^2}{\hbar^{2}}\frac{|\Omega|}{|k|}B^{z}\sigma^{z} = {\cal E}\sigma^{z}, \eeq
where ${\cal E} = \frac{2im^2}{\hbar^{2}}\frac{|\Omega|}{|k|}B^{z}.$
Or we can write \beq {\cal E}^{\sigma}_{i} = \mp \epsilon_{ijz}\frac{2im^2}{\hbar^{2}}\frac{|\Omega|}{|k|}B^{z}.\eeq
Here $\mp$ sign corresponds to the electron having spins parallel and anti-parallel to $z$ axis. Thus this effective spin dependent electric field, results in a spin
separation along $z$ direction. This generates a spin current.
%The in plane components of spin electric field are given by
%\beq {\cal E}^{\sigma}_{x} = 2\frac{\partial |\Omega|}{\partial t}\sigma_{x} + \frac{|\Omega|}{|k|}B^{z}\sigma_{y}\label{la},\eeq and
%  \beq {\cal E}^{\sigma}_{y} \propto - 2\frac{\partial \Omega}{\partial t}\sigma_{y} +\frac{|\Omega(t)|}{|k|}B^{z}\sigma_{x} \label{lbc}.\eeq
Obviously the spin dependent part of the electric field i.e the Yang mills electric field is proportional to the absolute value of the rotation frequency and the applied magnetic field $B_{z}.$
That is we can control the motion of spins by controlling the rotation. %Another interesting result that appears in our calculation is that, not only in the spin dependent part, the spin independent part i.e the charge contribution of the motive
%force also contains the effect of mechanical rotation via the term $\phi^{'}_{\Omega}$ in (\ref{va}).
  %Here $i = {x,y}$ and $+$ and $-$ are for $x$ and $y$ directed electric field respectively.
The spin independent part of this electric field
\beq {\cal E}_{0} = \frac{cm^2}{e\hbar |k|}\frac{\partial |\Omega|}{\partial t} .\eeq
This spin independent part of the electric field depends on the time variation of rotational frequency.

%\section{current due to spin electric field}
The spin dependent electric field can generate spin currents \cite{tan}. The total current of the system can be written as a product of spin conductance and the corresponding emf. as
\beq j^{\sigma}_{i} = \vec{\zeta} . \vec{\cal{E}^{'}} = [\zeta_{0}I +\zeta_{s}(\rho .\sigma)].[{\cal E}_{0}I + {\cal E}^{\sigma}_{i})],\eeq
where $\vec{\rho}$ is the spin polarization axis and  $\vec{\zeta}_{s}$ is the spin conductance.
%In the above calculations we do not consider the Stern Gerlach like contribution.
Thus assuming the polarization axis is along $z$ direction, the total current can be written as
\begin{eqnarray}\label{sc}
j_{x} &=& \left[\zeta_{0}I + \zeta_{s}\sigma^{z}].[{\cal E}_{0}I + ({-\cal E})\sigma^{z}\right]\nonumber\\
%&=& ( - \zeta_{s}\vec{\cal{E}}^{\sigma}_{x})I + (\zeta_{s}\partial_{x}\phi^{'}_{\Omega} - \zeta_{0}\vec{\cal{E}}^{\sigma}_{x} )\sigma^{y}\nonumber\\
&=& \left(\zeta_{0}\vec{\cal{E}}_{0} - \zeta_{s}{\cal E}\right)I + \left(\zeta_{s}\vec{\cal{E}}_{0} - \zeta_{0}{\cal E}\right)\sigma^{z}\nonumber\\
&=& \left(\zeta_{0}\frac{cm^2}{e\hbar |k|}\frac{\partial |\Omega|}{\partial t}- \zeta_{s}\frac{2im^2}{\hbar^{2}}\frac{|\Omega|}{|k|}B^{z}\right)I + \left(\zeta_{s}\frac{cm^2}{e\hbar |k|}\frac{\partial |\Omega|}{\partial t}- \zeta_{0}\frac{2im^2}{\hbar^{2}}\frac{|\Omega|}{|k|}B^{z}\right)\sigma^{z}\nonumber\\
&=& j_{x}^{c}I + j^{s}_{x}\sigma^{z}
\end{eqnarray}
where
\begin{eqnarray}
j_{x}^{c} = \zeta_{0}\frac{m^2}{e\hbar |k|}\frac{\partial |\Omega|}{\partial t}- \zeta_{s}\frac{2im^2}{\hbar^{2}}\frac{|\Omega|}{|k|}B^{z}
\end{eqnarray}
is the charge current
 and
\begin{eqnarray}
j^{s}_{x} = \zeta_{s}\frac{cm^2}{e\hbar |k|}\frac{\partial |\Omega|}{\partial t}- \zeta_{0}\frac{2im^2}{\hbar^{2}}\frac{|\Omega|}{|k|}B^{z}
\end{eqnarray}
is the spin current of the system in the $x$ direction and it varies with mechanical rotation.
If we switch off the rotation, the spin current as well as charge current vanishes. Importantly, the term $\zeta_{s}{\cal E}$ drives the charge current, whereas
$\zeta_{s}{\cal E}_{0}$ drives the spin current. The spin dependent electric field generates as a consequence of spin rotation coupling induced SOC and can induce
not only a spin current, but also a charge current in the system. This effect is known as inverse spin Hall effect \cite{shibata}.
%Thus we can see that the charge current as well as the spin current contains the effect of rotation through the $\phi^{'}$ term, which imply that in presence of external fields the inertial potential modifies both the spin current and charge current.

\section{renormalization effect on spin dependent electric field, spin current and Berry curvature}
The spin dynamics of the semiconductors are influenced by the $\vec{k} . \vec{p}$ perturbation theory \cite{winkler} as the band structure of a semiconductor in the vicinity of the band edges can be very well described by the $\vec{ k} . \vec{ p} $ method.
On the basis of $\vec{ k} . \vec{ p}$ perturbation theory, by taking into account the interband mixing, one can explain many characteristic features of semiconducting system .
It is well known that due to $\vec{k}.\vec{p}$ perturbation, the spin orbit coupling as well as Zeeman coupling get renormalized. In a rotating as well as accelerated frame the conventional $\vec{k}.\vec{p}$ method can be generalized with the modified momentum $k_{\Omega}$, where $\vec{\pi}_{\Omega} = \vec{p} - e\vec{A}^{'} = \hbar \vec{k}_{\Omega},$ where $\vec{A}^{'} = \vec{A} + \vec{A}_{\Omega}$. The $\vec{k}.\vec{p}$ perturbed Hamiltonian for the inertial system \cite{winkler} is given as
\begin{eqnarray}
H_{kp} &=& \frac{P^2}{3}\left(\frac{2}{E_{G}} + \frac{1}{E_{G} + \triangle_{0}}\right)\vec{k}_{\Omega}^{2}  - \frac{P^2}{3}\left(\frac{1}{E_{G}} - \frac{1}{(E_{G} + \triangle_{0})}\right)\frac{e}{\hbar}\vec{\sigma}.\vec{B}\nonumber\\ &-& \frac{P^2}{3}\left(\frac{1}{E_{G}} - \frac{1}{(E_{G} + \triangle_{0})}\right)\frac{e}{\hbar}\vec{\sigma}.\vec{B}_{\Omega} - e\frac{P^2}{3}\left(\frac{1}{E_{G}^{2}} - \frac{1}{(E_{G} + \triangle_{0})^{2}}\right)\vec{\sigma}.(\vec{k}_{\Omega}\times \vec{E}_{a})\label{kp}
\end{eqnarray}
Adding Hamiltonian (\ref{kp}) and (5) we get
the total Hamiltonian for the inertial system within the $\vec{k} . \vec{p}$ formulation as \cite{matsuo kp, bc}
\beq H_{\Omega, kp} = \frac{\vec{\pi}_{\Omega}^{2}}{2m^*} - \phi_{I} - (1 + \frac{\delta g}{2})\mu\vec{\sigma} . \vec{B}  - (1 + \delta g)\mu\vec{\sigma} . \frac{\vec{B}_{\Omega}}{2} - e(\lambda + \delta \lambda)\vec{ \sigma} .(\vec{ k}_{\Omega}\times \vec{ E}_{a}) , \label{f}\eeq
where $\phi_{I} = e(V_{a} + \phi_{\Omega}),$ the total potential of the system and $ \lambda = \frac{\hbar^{2}}{4m^{2}c^{2}}$ is the spin orbit coupling strength as considered in  vacuum. Furthermore, the perturbation parameters $\delta g$ and $\delta \lambda$ are given by
\begin{eqnarray}\label{lam}
\delta g &=& -\frac{4m}{\hbar^{2}}\frac{P^2}{3}\left(\frac{1}{E_{G}} - \frac{1}{E_{G} + \triangle_{0}}\right)\nonumber\\
\delta \lambda &=& + \frac{P^2}{3}\left(\frac{1}{E_{G}^{2}} - \frac{1}{(E_{G} + \triangle_{0})^{2}}\right)
\end{eqnarray} are the renormalization parameter for the Zeeman coupling term and spin orbit coupling terms respectively. $\frac{1}{m^*} = \frac{1}{m} + \frac{2P^2}{3\hbar^{2}}\left(\frac{2}{E_{G}} + \frac{1}{E_{G} + \triangle_{0}}\right)$ is the effective mass term. Here $P, E_{G}, \triangle_{0}$ are the Kane momentum matrix element, band gap energy and the spin orbit gap parameter respectively, together known as the Kane model parameters \cite{winkler}. Due to this renormalization, we try to find out the renormalization of the rotation induced spin dependent electric field and spin current.
In order to calculate that, we write the Hamiltonian, considering only the rotation induced terms as
\beq H_{\Omega, kp} = \frac{1}{2m^*}\left(\vec{p}- \frac{e}{c}\vec{A}^{'}- \frac{\mu}{c}A_{\sigma, kp}\right)^{2} - e\phi_{\Omega} - (1 + \frac{\delta g}{2})\mu\vec{\sigma} . \vec{B}\label{cv},\eeq
%\beq H_{\Omega, kp} = \frac{1}{2m^*}\left(\vec{\pi}_{\Omega} - {\cal \vec{A}}_{\sigma, kp}\right)^{2} + e\phi^{'}_{\Omega} - (1 + \frac{\delta g}{2})\mu_{B}\sigma^{z}B^{z}.\label{cl}\eeq
Here the corresponding gauge is ${\cal A}_{\sigma, kp} = (1 + \delta g)\frac{(m^*)^2c^{2}|\Omega|}{e\hbar|k|}(\sigma_{x} , - \sigma_{y}, 0),$ which suggests the renormalization of
the spin dependent gauge. Following the methodology used in the previous section, we can evaluate the $x$ component of this electric field  for the modified SU(2) gauge  $a^{\mu} = (a^{0}_{kp}, \vec{a}_{kp}) = (-(1 + \frac{\delta g}{2})\vec{\sigma} . \vec{B}, {\cal A}_{\sigma, kp})$
 as
%\beq {\cal E}_{x, kp} =  (1 + \frac{\delta g}{2})(1 + \delta g)\frac{i(m^*)^2}{\hbar^{3}}\frac{|\Omega|}{|k|}B^{z}\left[\sigma^{x},\sigma^{z}\right].\label{ca}\eeq
%Using the commutation relation between the Pauli matrices we can rewrite (\ref{ca})  as
\beq {\cal E}_{x, kp}\sigma^{x} =  -(1 + \delta g)\frac{c(m^*)^2}{e\hbar |k|}\frac{\partial|\Omega|}{\partial t} -(1 + \frac{\delta g}{2})(1 + \delta g)\frac{2i(m^*)^2}{\hbar^{2}}\frac{|\Omega|}{|k|}B^{z}\sigma^{z}\label{va}. \eeq
The $x$ component of the spin dependent part of the electric field is
\beq {\cal E}_{x, kp}^{\sigma} =  -(1 + \frac{\delta g}{2})(1 + \delta g)\frac{2i(m^*)^2}{\hbar^{2}}\frac{|\Omega|}{|k|}B^{z}\sigma^{z}\label{va} \eeq
and similarly the $y$ component of this spin electric field is
\beq {\cal E}_{y, kp}^{\sigma} =   (1 + \frac{\delta g}{2})(1 + \delta g)\frac{2i(m^*)^2}{\hbar^{2}}\frac{|\Omega|}{|k|}B^{z}\sigma^{z}. \eeq
The spin contribution of this electric field in $x$ and $y$ direction are given by
\beq {\cal E}^{\sigma}_{x, kp} = - (1 + \frac{\delta g}{2})(1 + \delta g)\frac{2i(m^*)^2}{\hbar^{2}}\frac{|\Omega|}{|k|}B^{z}\sigma^{z} = -{\cal E}_{kp}\sigma^{z}\label{lac},\eeq and
  \beq {\cal E}^{\sigma}_{y, kp} = (1 + \frac{\delta g}{2})(1 + \delta g)\frac{2i(m^*)^2}{\hbar^{2}}\frac{|\Omega|}{|k|}B^{z}\sigma^{z}\label{la}.\eeq
Upto first order of $\delta g$ we can write the $x$ component of the spin electric field as
\beq {\cal E}^{\sigma}_{x, kp} = - (1 + \frac{3}{2} \delta g)\frac{2i(m^*)^2}{\hbar^{2}}\frac{|\Omega|}{|k|}B^{z}\sigma^{z}\label{la}.\eeq
In tensorial form we can write
\beq {\cal E}^{\sigma}_{i, kp} = \mp \epsilon_{ijz}(1 + \frac{3}{2} \delta g)\frac{2i(m^*)^2}{\hbar^{2}}\frac{|\Omega|}{|k|}B^{z}.\eeq
The rotation induced spin electric field is then renormalized by a factor $(1 + \frac{3}{2} \delta g).$ For a solid, having the spin orbit gap parameter equals to zero, we get no contribution for the $\vec{k} . \vec{p}$ perturbation. The ratio of the spin dependent electric field with and without the $\vec{k} . \vec{p}$ perturbation can be obtained as
\beq |\frac{{\cal E}^{\sigma}_{i, kp}}{{\cal E}^{\sigma}_{i}}| \propto  (1 + \frac{3}{2} \delta g)\label{la},\eeq
where $i = {x, y}.$ Here it is evident from our result that similar to the renormalization of SRC and Zeeman term, spin electric field is also renormalized. But the prefactor is different from the two types of coupling. This is an important result of our work.

Due to the application of $\vec{k} . \vec{p}$ perturbation, the spin current is also renormalized. The expression of the total current can be readily obtained from (\ref{sc}) as
\begin{eqnarray}\label{scu}
j_{x, kp} &=& \left[\zeta_{0}I + \zeta_{s}\sigma^{z}].[{\cal E}_{0, kp}I + ({-\cal E}_{kp}\sigma^{z})\right]\nonumber\\
&=& \left(\zeta_{0}\vec{\cal{E}}_{0, kp} - \zeta_{s}{\cal E}_{kp}\right)I + \left(\zeta_{s}\vec{\cal{E}}_{0, kp} - \zeta_{0}{\cal E}\right)\sigma^{z}\nonumber\\
&=& \left(\zeta_{0}(1 + \delta g)\frac{cm^2}{e\hbar |k|}\frac{\partial |\Omega|}{\partial t}- \zeta_{s}(1 + \frac{3}{2} \delta g)\frac{2im^2}{\hbar^{2}}\frac{|\Omega|}{|k|}B^{z}\right)I + \left(\zeta_{s}(1 + \delta g)\frac{cm^2}{e\hbar |k|}\frac{\partial |\Omega|}{\partial t}- \zeta_{0}(1 + \frac{3}{2} \delta g)\frac{2im^2}{\hbar^{2}}\frac{|\Omega|}{|k|}B^{z}\right)\sigma^{z}\nonumber\\
&=& j_{x, kp}^{c}I + j^{s}_{x, kp}\sigma^{z}
%&=& (1 + \delta g)j_{x}^{c}I + (1 + \frac{3}{2} \delta g)j^{s}_{x}\sigma^{z},
\end{eqnarray}
where \beq j_{x, kp}^{c} = \zeta_{0}(1 + \delta g)\frac{cm^2}{e\hbar |k|}\frac{\partial |\Omega|}{\partial t}-
\zeta_{s}(1 + \frac{3}{2} \delta g)\frac{2im^2}{\hbar^{2}}\frac{|\Omega|}{|k|}B^{z}\eeq and \beq j^{s}_{x, kp} = \zeta_{s}(1 + \delta g)\frac{cm^2}{e\hbar |k|}\frac{\partial |\Omega|}{\partial t}-
\zeta_{0}(1 + \frac{3}{2} \delta g)\frac{2im^2}{\hbar^{2}}\frac{|\Omega|}{|k|}B^{z}\eeq are the modified charge and spin current in the $x$ direction i.e the charge
and spin current are renormalized.

 %Interestingly the spin current is renormalized by the factor $(1 + \frac{3}{2}\delta g)$.

Next we investigate the renormalization of Berry curvature in presence of $\vec{k} . \vec{p}$ method
and the expression of the Berry curvature can be evaluated as
%\beq M_{z, kp} = -\frac{2(m^*)^4}{e\hbar^{5}}\frac{|\Omega|^{2}}{|k|^{2}}(1 + 2\delta g + (\delta g)^{2})\sigma^{z}.\eeq
\beq M_{z, kp} = -\frac{2c^{3}(m^*)^4}{e\hbar^{3}}\frac{|\Omega|^{2}}{|k|^2}(1 + 2\delta g)\sigma^{z},\eeq
where the terms upto first order  of $\delta g,$  are retained.
Then the ratio of the Berry curvature, with and without $\vec{k} . \vec{p}$ method is given as
\beq \frac{M_{z, kp}}{M_{z}} \propto  (1 + 2\delta g).\eeq
Thus we noticed that due to $\vec{k} . \vec{p}$ perturbation the spin electric field, spin current and Berry curvature are renormalized differently. This shows that the spatial and temporal parts of the field tensor are modified differently due to $\vec{k} . \vec{p}$ perturbation. In \cite{matsuo kp}, the renormalization of SRC is discussed, which is proportional to $(1 + \delta g)$, whereas that for the spin electric field is $(1 + \frac{3}{2}\delta g)$ and Berry curvature is $(1 + 2\delta g).$ This is one of our main result. Importantly, the spin dependent electric field, spin current and Berry curvature depends on the mechanical rotation. Thus by controlling the rotation frequency, one can control the spin electric field, spin current externally. This could be very useful in the area of spintronic applications.

\section{Inertial effect on spin splitting, spin Berry phase, spin galvanic effect}
%Spin  is normally controlled by magnetic field, however
%in semiconductors the spin can also be controlled by spin-orbit interaction, the strength of which in semiconductor nanostructures can be tuned by an external field. In spintronics the preference is given to electric fields rather than magnetic fields.
To discuss the spin splitting scenario, we start from the Hamiltonian in both accelerated and rotated frame of reference.
The Hamiltonian in presence of the induced electric  and magnetic fields due to inertial effect can be written as
\beq H = \frac{\vec{p}^2}{2m} - eV_{a}(\vec{r}) -\vec{\Omega}.(\vec{r}\times \vec{p})- \frac{e\hbar}{4m^{2}c^{2}}\vec{\sigma}.\left(\vec{p} \times \vec{E}_{a}\right) - \frac{e\hbar}{4m}\vec{\sigma}.\vec{B}_{\Omega} ,\label{hh} \eeq where the third term in the r.h.s of Hamiltonian (\ref{hc}) is the coupling term between rotation and orbital angular momentum. The gauge potential $\vec{A}_{\Omega}$ originates from this coupling term. On the other hand the Zeeman like coupling (last term in the r.h.s of (\ref{hh}))
is due to the spin rotation coupling. The forth term in the r.h.s of (\ref{hh}) is the inertial spin orbit coupling term. The acceleration induced
electric field has important contribution on the spin current and spin polarization which is explained in reference \cite{bc, cb}.
%In the Hamiltonian (\ref{hc}), the last term is zeeman like due to the spin rotation coupling term.
Neglecting the terms with order greater than $\frac{1}{c^{2}},$ the Hamiltonian in (\ref{hh}) can be rewritten in the following form
\beq H = \frac{1}{2m}\left(\vec{p} - \frac{e}{c}\vec{A}_{\Omega} - \frac{\mu}{2c}\vec{E}_{a}\times \vec{\sigma}\right)^{2} - \phi_{I} -
\frac{\mu}{2}\sigma . \vec{B}_{\Omega} ,\label{ab}\eeq where $\mu = \frac{e\hbar}{2mc}$ is the magnetic moment of electron and $\phi_{I} = e(V_{a} + \phi_{\Omega}),$ is the total inertial potential of the system and $\phi_{\Omega} = \frac{m}{2e}(\vec{\Omega}\times\vec{r})^{2}$ is the
induced potential due to rotation and $\vec{A}_{\Omega}$ is the gauge of this induced magnetic field.

We first try to find out the relation between the spin rotation coupling with the Berry phase \cite{berry}. Next we theoretically develop the idea of spin galvanic effect in an inertial frame and try to find out which contributions in the spin splitting is favorable, acceleration or rotation.

\subsection{Inertial SOC , spin splitting and Berry phase}
In the presence of the magnetic field $\vec{B}$, the Zeeman term splits the spin of degenerate energy  $\epsilon_{\uparrow}$ and $\epsilon_{\downarrow}$, the spins of which are pointing in
opposite directions with respect to the direction of the magnetic field. The difference of energy $\triangle \epsilon = \epsilon_{\uparrow} - \epsilon_{\downarrow}$ is usually proportional to the magnetic field strength and is known as the Zeeman spin splitting energy.
%Now The Rashba SOC term ensures that one can still get  spin split parts when the external magnetic fields are zero.
%The microscopic
%origin of terms linear in electron wavevector
%in low dimensional systems is structure inversion asymmetry
%(SIA) and bulk inversion asymmetry (BIA) which
%lead to Rashba and Dresselhaus spin-orbit terms in the
%Hamiltonian, respectively.
In the previous sections, we have achieved two different coupling terms, similar to the Rashba and Dresselhaus terms \cite{winkler}. The Hamiltonian in terms of this two different coupling due to inertial effects when we consider the electric field is along the $z$ direction can be written as
\beq H = \frac{\hbar^{2}\vec{k}_{\Omega}^2}{2m} + \alpha_{a}(k_{x, \Omega}\sigma_{y} - k_{y, \Omega}\sigma_{x}) + \beta_{\Omega} (\sigma_{y}k_{y,\Omega} - \sigma_{x}k_{x,\Omega}) ,\label{as}\eeq
where the $\alpha_{a} = \lambda E_{a, z},$ is the Rashba like coupling parameter appears in an accelerating system.
The knowledge of the relative strength of Rashba and Dresselhaus like terms is important for investigations of different spin dependent phenomena
in our inertial system. Interestingly we found that unlike to the original Dresselhaus effect, in our system we can externally control the Dresselhaus like spin splitting through rotation. Thus in our inertial system we can produce a large spin splitting by increasing acceleration and rotation externally.

%\underline{\textbf{Case $1$: $\vec{a} \ne 0, \Omega \ne 0 , \vec{E} = 0$}} \\
%If both the magnetic fields due to acceleration and rotation in $\vec{k}$ space is present, then we can explain the splitting with the help of the effective magnetic field $\vec{B}_{I} = \vec{B}_{\vec{a}} + \frac{e\hbar}{2m\gamma}\vec{B}_{\vec{\Omega}}.$
%%Thus we can show the splitting is modified in presence of both acceleration and rotation.
%In presence of both terms, the spin splitting gets modified.
%\underline{\textbf{Case $2$: $\vec{a} \ne 0, \Omega = 0, \vec{E} = 0$}} \\
 %The dispersion  relation for $\Omega = 0$ is given as
 The dispersion relation for the Hamiltonian in (\ref{as}) can be obtained as
\beq \epsilon_{\pm}(\vec{k}_{\pm}) =  \frac{\hbar^{2}\vec{k}_{\pm}^2}{2m}  \pm \vec{k}_{\pm}  \sqrt{\alpha_{a}^{2} + \beta_{\Omega}^{2} + 2\alpha_{a}\beta_{\Omega}cos2\phi},\label{epsi}\eeq
where $\vec{k}_{\pm} = k_{x} \pm ik_{y}.$
The dispersion relation can be rewritten as
\beq \epsilon_{\pm}(\vec{k}_{\pm}) =  \frac{\hbar^{2}\vec{k}_{\pm}^2}{2m}  \pm k_{\pm}\eta(a,\Omega,\phi)\label{eps},\eeq
where $\phi$ is the angle between $k$ with the $x$ axis and $\eta(a,\Omega,\phi) = \sqrt{\alpha_{a}^{2}+ \beta_{\Omega}^{2} + 2\alpha_{a}\beta_{\Omega} cos2\phi}.$ The two bands only meet at $k = 0$ point. The difference between the energy of up branch and down branch is known as the spin split energy and given by
\beq \epsilon_{+} - \epsilon_{-} = \frac{2m}{\hbar^{2}}\eta (a,\Omega,\phi), \eeq
which is independent of $\vec{k}$.
For small $\vec{k}$ the second term in (\ref{eps}) is the dominated term. Usually in Rashba coupling one get a ring of minima for the brunch $\epsilon_{-}(\vec{k}_{\pm})$.
 In our case we can write the minimum value as
\beq \epsilon_{-}(\vec{k}_{\pm}) = -\frac{ m \eta^{2}(a,\Omega,\phi)}{2\hbar^2} .\eeq
Here the minima occurs at \beq  k_{m} = \frac{m\eta(a,\Omega,\phi)}{\hbar^2}\eeq
The quantum density of spin can be given by
\beq N_{m} = \frac{(2k_{m})^{2}}{4\pi}  = \frac{m^{2}}{\pi\hbar^4} \eta^{2}(a,\Omega,\phi).\eeq
Now if the total density of electron is greater than the quantum density we only have the lower sub-band occupied. The total 2D density of the system $N_{s}$ can be written as the sum of the unequal populations $N_{\pm}$ as
\begin{eqnarray}
N_{s} &=& N_{+} + N_{-} \\
N_{\pm} &=& \frac{k_{F \pm}^{2}}{4\pi} ,
\end{eqnarray}
where $k_{F \pm}$ are the Fermi wave vectors of the spin split branches. Now the Fermi energy in terms of total density $N_{s}$ can be written as
\begin{equation}
E_{F} = \frac{\pi N_{s}\hbar^2}{m} - \frac{m(\eta^{2}(a,\Omega,\phi))}{\hbar^2}
\end{equation}
Thus we can give the expression of the population of  the spin-split brunch as
\begin{equation}
N_{\pm} = \frac{N_{s}}{2} \mp \frac{(\eta(a,\Omega,\phi))m^2}{2\hbar^4\pi}\sqrt{\frac{2\pi\hbar^4}{m^{2}}N_{s} - (\eta(a,\Omega,\phi))^{2} }
\end{equation}
From the above eqns we can find out the value of $\eta(a,\Omega,\phi)$ as
\beq \eta(a,\Omega,\phi) = \frac{\hbar^{2}}{2m}(k_{F_{+}} - k_{F_{-}}) .\eeq
Lastly the density of states can be obtained for our inertial system  when $E \geq 0$ as
\beq D_{\pm} = \frac{m}{2\pi\hbar^2}\left(1 \mp \frac{\eta(a,\Omega,\phi)}{\sqrt{\frac{2\pi\hbar^2}{m}}E + \eta^{2}(a,\Omega,\phi)}\right),\eeq
and for $E < 0$ we have
\beq D_{-} = \frac{m}{\pi\hbar^2}\left( \frac{\eta(a,\Omega,\phi)}{\sqrt{\frac{2\pi\hbar^2}{m}}E + \eta^{2}(a,\Omega,\phi)}\right).\eeq
Thus we can easily see that the density of state becomes singular when $E = \epsilon_{-}$, i.e the $Van Hove$ singularity can also be achieved for our inertial system, without the application of the external fields.
The eigenstates of Hamiltonian (\ref{ab}) will also be the eigenstate of wave vector $\vec{k}$ and is given as
\begin{eqnarray} \label{ket}
|k_{+}, +, \theta\rangle = \frac{1}{\sqrt{2}}\left( \begin{array}{c}
e^{-i\theta}\\
+i \end{array} \right)\otimes|k_{+}\rangle, ~~~~~~~
|k_{-}, -, \theta\rangle = \frac{1}{\sqrt{2}}\left( \begin{array}{c}
e^{-i\theta}\\
-i \end{array} \right)\otimes|k_{-}\rangle ,
\end{eqnarray}
where $|k\rangle$ is the eigenstate of $\vec{k}$ and the parameter $\theta$ is given by
\beq tan\theta = \frac{\alpha_{a}k_{y} - \beta_{\Omega}k_{x}}{\alpha_{a}k_{x} - \beta_{\Omega}k_{y}}.\eeq
 With the help of the eigenkets in equn. (\ref{ket}), we can find out the Berry phase for this inertial system. Thus the spin Berry phase in terms of the Rashba and Dresselhaus like coupling parameter can be written as
\beq \gamma_{\pm} = \oint dl.\left\langle k_{\pm}, \pm, \theta|i\frac{\partial}{\partial k}|k_{\pm}, \pm, \theta\right\rangle = \frac{\alpha_{a}^{2} - \beta_{\Omega}^{2}}{|\alpha_{a}^{2} - \beta_{\Omega}^{2}|}\pi.\label{x}\eeq
The Berry phase \cite{berry} plays a very important role in electron transport. It is clear from (\ref{x}) that the Berry phase is zero for $|\alpha_{a}| = |\beta_{\Omega}|$ i.e when $\frac{\hbar^{2}}{4emc^2}|a_{z}| =  \frac{|\Omega|}{2|k|}$ or $|a_{z}| = \frac{2emc^2}{\hbar^{2}|k|}|\Omega|.$ The vanishing Berry phase at $|a_{z}| = \frac{2mec^2}{\hbar^{2} |k|}|\Omega|$ is relevant to the crossover of two energy bands.  When $|a_{z}| \neq \frac{2mec^2}{\hbar^{2}|k|}|\Omega|$ we have non vanishing Berry phase.
In this regard another important aspect is obtained when we add an external electric field with vanishing rotation. In this case the Hamiltonian can be obtained as
\beq H = \frac{\hbar^{2}\vec{k}^2}{2m} + (\alpha - \alpha_{a})( k_{y}\sigma_{x} - k_{x}\sigma_{y}) .\label{rasba}\eeq
For a critical value of acceleration i.e when we have $\frac{m}{e}a_{z} = E_{z},$ we can show that the ring of minima converge to a point. We call this acceleration as the critical acceleration.
Thus in presence of both the acceleration and external electric field, at critical acceleration we don't get any spin split part at $\Omega = 0.$ But if we have both the rotation term and external magnetic field, the electric field acting on the particle gets modified due to the presence of rotation \cite{matsuoprb}.

The Rashba like and Dresselhaus like spin splitting due to the acceleration and rotation, results different patterns in the $\vec{k}$ space. Next we try to distinguish the relative contributions of acceleration and rotation induced SO coupling. This can be done by the spin galvanic effect \cite{Ganichev}.

%The Berry phase is related to the the $\vec{k}$ space spin Hall conductance via the following relation \cite{sqshen}

\subsection{Inertial spin galvanic effect}
The spin galvanic effect(SGE) is due to the non-equilibrium but uniform population of electron spin. This non-equilibrium spin injection can be done optically or by non-optical means. The origin of this effect is the shift of spin-up and spin down electrons in the momentum space and the spin flip scattering events between the two sub-bands. In this case it is possible to measure the anisotropic orientation of spins in the momentum space and hence the different
contribution of the Rashba and the Dresselhaus terms \cite{Ganichev1}.
%A method of measuring the relative effects of the Rashba and Dresselhaus strengths can be done with the
%photo current experiment.
The spin galvanic effect states that as the
asymmetric spin levels are split, with one spin orientation favored, an excitation can
cause electrons to reverse spin, creating a charge current along the x-y plane. The direction of
this charge current depends on the spin polarization within the lattice.
From the Hamiltonian (\ref{sss}), the total current can be determined as a function of the spin
polarization along the surface.
The $SGE$ current $j_{SGE}$ and
the average spin are related by a second rank pseudo-tensor
with components proportional to the parameters
of spin-orbit splitting as follows \cite{Ganichev1}
\begin{equation}\label{alf}
j_{SGE} = A\left( \begin{array}{cc}
\beta_{\Omega} & -\alpha_{a} \\
  \alpha_{a} & \beta_{\Omega}
\end{array} \right)S
\end{equation}
where $S$ is the average spin in the plane and $A$ is a constant determined by the kinetics
of the $SGE$, namely by the characteristics of momentum
and spin relaxation processes. Unlike the usual cases, we have here two inertial spin orbit coupling strengths coming as a consequence of SOC terms due to acceleration and the spin rotation coupling effect. The spin galvanic current in (\ref{alf}) can be rewritten in terms of the acceleration and rotation as
\begin{equation}\label{al}
j_{SGE} = A\left( \begin{array}{cc}
\frac{|\Omega|}{2|k|} & - \frac{\hbar^{2}}{4mec^2}a_{z} \\
  \frac{\hbar^{2}}{4mec^2}a_{z} & \frac{|\Omega|}{2|k|}
\end{array} \right)S
\end{equation}
Eqn. (\ref{al}) helps one to find out the ratio of $\alpha_{a}$ and $\beta_{\Omega}$ or the ratio of acceleration and rotation. The spin galvanic current can be decomposed into the current duo to Rashba like term and Dresselhaus like term $j_{a}$ and $\vec{j}_{\Omega}$ proportional to the $\alpha_{a}$ and $\beta_{\Omega}$ term. From symmetry, $\vec{j}_{a}$ is perpendicular to average spin $\vec{S}$, whereas $\vec{j}_{\Omega}$  makes an angle $2\Psi$ with $\vec{S}$, where $\Psi$ is the angle between $\vec{S}$ and the $x$ axis.
Thus one can find out the spin galvanic currents in terms of the coupling terms due to acceleration and rotation. The absolute value of the total current $j_{SGE}$
is given by the expression
\beq j_{SGE} = \sqrt{j_{a}^{2} + j_{\Omega}^{2} - 2j_{a}j_{\Omega}sin 2\Psi}\label{m}\eeq
Now suppose the average spin vector is oriented along the $x$ axis, then we have angle $\Psi = 0$. Thus from (\ref{m}), we have $\vec{j}_{a}$ and $j_{\Omega}$ are directed perpendicular and parallel to $\vec{S}$, respectively. The ratio of the currents in the $x$ and $y$ direction is  can be obtained
\beq \frac{j_{y}}{j_{x}} = \frac{\alpha_{a}}{\beta_{\Omega}} ,\eeq
or interestingly
\beq \frac{a_{z}}{|\Omega|} \propto \frac{j_{y}}{j_{x}}.\eeq
The above ratio gives us the ratio of acceleration to rotation. Thus controlling the two inertial parameters we can control the charge current flowing through the system. Interestingly, this ratio is an important ingredient to understand whether acceleration or rotation contribution is dominating. We should mention here that for different orientation of $\vec{S}$ we get different value of this ratio. Thus without any external fields we are successful to produce charge current from the $k$ linear terms of the Hamiltonian, due to inertial effects.  %The maximum and minimum spin splitting of the two types can be obtained as $2(\alpha_{a} + \beta_{\Omega})k$ and $2(\alpha_{a} - \beta_{\Omega})k $ respectively. On the other hand the ratio of currents in the $x$ and $y$ direction can be controlled by the acceleration and rotation. Interestingly one can control the ratio of the current due to acceleration or rotation term by controlling the ratio of acceleration and rotation.

\section{conclusion}
In the present paper, we theoretically derive the expression of spin rotation coupling induced spin dependent electric field in a rotating system. SRC plays a very important role in achieving a SOC like coupling, from which we can find out the spin electric field for our system. Furthermore, from the rotation induced SOC term, one can achieve the expression of spin current and Berry curvature for the rotating system.
In addition to that, by taking into account the interband mixing, using the eight band Kane model we study the renormalization of the spin dependent electric field, spin current and Berry curvature. Due to the application of $\vec{k}. \vec{p}$ method, where the spin electric field is modified by the factor $(1 + \frac{3}{2}\delta g ),$ the Berry curvature modifies as $(1 + 2\delta g).$ Interestingly, we find that the spin electric field, Berry curvature as well as spin current are renormalized differently from the well known coupling terms(Zeeman, SRC) due to presence of different prefactor of $\delta g.$ Furthermore, the effect of rotation on Berry phase as well as spin splitting is investigated. %One can observe that there exists similar kind of spin splitting in a semiconductor as Rashba and Dresselhaus type.
One can also explain inertial spin galvanic effect from the interplay of acceleration and rotation induced coupling constant. The ratio of the acceleration induced Rashba like and rotation induced Dresselhaus like coupling can be obtained as well.

\end{document}